\documentclass[prl,twocolumn,showpacs,superscriptaddress]{revtex4}
\usepackage{amsmath,amsfonts,paralist}
\usepackage{graphicx}

\newcommand{\DL}{$D_{\rm L}$}
\newcommand{\tC}{\widetilde C}
\newcommand{\chiSG}{\chi_{\rm sg}}

\begin{document}

\title{Ising spin glass transition in magnetic field out of mean-field}

\author{L. Leuzzi} %\email{luca.leuzzi@roma1.infn.it}
\affiliation{Dipartimento di Fisica, Sapienza Universit\`a di Roma,
  P.le Aldo Moro 2, I-00185 Roma, Italy}
\affiliation{Statistical Mechanics and Complexity Center (SMC) - INFM
  - CNR, Italy}

\author{G. Parisi} %\email{giorgio.parisi@roma1.infn.it}
\affiliation{Dipartimento di Fisica, Sapienza Universit\`a di Roma,
  P.le Aldo Moro 2, I-00185 Roma, Italy}
\affiliation{Statistical Mechanics and Complexity Center (SMC) - INFM
  - CNR, Italy}

\author{F. Ricci-Tersenghi} %\email{federico.ricci@roma1.infn.it}
\affiliation{Dipartimento di Fisica, Sapienza Universit\`a di Roma,
  P.le Aldo Moro 2, I-00185 Roma, Italy}

\author{J.J. Ruiz-Lorenzo} %\email{ruiz@unex.es}
\affiliation{Departamento de F\'{\i}sica, Univ. Extremadura, Badajoz,
  E-06071 and BIFI, Spain.}

\begin{abstract}

The spin-glass transition in external magnetic field is studied both
in and out of the limit of validity of mean-field theory on a diluted
one dimensional chain of Ising spins where exchange bonds occur with a
probability decaying as the inverse power of the distance.  Varying
the power in this long-range model corresponds, in a one-to-one
relationship, to change the dimension in spin-glass short-range
models.  Evidence for a spin-glass transition in magnetic field is
found also for systems whose equivalent dimension is below the upper
critical dimension at zero magnetic field.

\end{abstract}

%{Spin-glass and other random models}
%{Disordered structures; amorphous and glassy solids}
%{Phase transitions: general studies}

\maketitle

{\em Introduction --}
Even though 30 years have passed since the spin-glass (SG) phase in
presence of an external magnetic field has been characterized in
mean-field theory \cite{Almeida78}, its existence in realistic
finite-dimensional systems is not yet an established issue.  In most
common (Heisenberg-like) amorphous magnets, e.g., AgMn, CuMn, and
AuFe, a SG phase has been detected also in presence of an external
field \cite{Petit}.  In mean-field theory of vectorial spin-glasses
this transition is expected along the so-called Gabay-Toulouse line
\cite{Gabay81}.
In Ising-like materials, instead, like Fe$_x$Mn$_{1-x}$TiO$_3$, it is
still a matter of debate whether or not a SG phase occurs when the
system is embedded in a magnetic field \cite{Petit,Jonsson05}.
Irreversible phenomena are, actually, detected in experiments as the
temperature is lowered: the separation of zero-field cooled and
field-cooled susceptibilities (or magnetizations) and the rapid
increase of characteristic relaxation times.  In zero field these are
the signatures of a thermodynamic transition, but in some recent AC
measurements in a magnetic field \cite{Jonsson05}, their magnitude
tends to depend sensitively on frequency and they are interpreted as
pertaining to a glassy dynamic arrest, rather than to a true
thermodynamic transition. According to this, the SG features measured
in a field would be artifacts of being out of equilibrium, similarly
to what happens in the structural glass transition, in which the
liquid glass former falls out of equilibrium at some low $T$ when its
structural relaxation time becomes longer than the observation time
and it vitrifies into an amorphous solid.

The Replica Symmetry Breaking (RSB) theory, holding in the mean-field
regime for spin-glasses, predicts a thermodynamic transition in
magnetic field $h$ at a finite temperature \cite{Parisi80}. In this
framework, a transition line, called Almeida-Thouless (AT)
\cite{Almeida78} line, can be identified in the $T-h$ plane between a
paramagnetic and a spin glass phase. At sufficiently low dimensions
(i.e. below the lower critical dimension, \DL) the transition
disappears. The value of \DL\ is not known, but it is quite possible
that in a field \DL\ is higher than for $h=0$ (as it happens for a
ferromagnet in a random field). There are some numerical evidences
(and analytic results) supporting \DL$=2.5$ at zero field.  For $h>0$,
some arguments suggested \DL=6, but recently Temesvari
\cite{Temesvari08} argued that the AT line can be continued below
$D=6$.
In the droplet theory, instead, no transition is predicted to remain
as soon as an infinitesimal field is switched on, independently from
the value of $D$. A crossover length $\ell_{d}(h,T)$ is introduced
\cite{FH88_386}, beyond which the SG phase is destroyed by the
field. The predictions of TNT scenario \cite{Krzakala00} should be
similar to those of the droplet model.
Extensive numerical works on the Edwards-Anderson model in 4D and 3D
yielded evidence both in favor of a transition in field
\cite{Marinari98,Krzakala01} and against it
\cite{Young04,Sasaki07,Jorg08}.
Unfortunately, finite size corrections are very strong in the presence
of an external field and it is hard to say whether these simulations
were really testing the thermodynamic limit.  To overcome this problem
we use a recently introduced SG model \cite{Leuzzi08}, which can be
simulated very efficiently, and a new data analysis, which should be
less sensitive to finite size effects.  We report numerical evidences
for a thermodynamic phase transition in the presence of external
fields also in systems for which the mean-field approximation is not
correct.

{\em The model --}
We investigate a one dimensional chain of $L$ Ising spins
($\sigma_i=\pm 1$) whose Hamiltonian reads \cite{Leuzzi08}
\begin{equation}
{\cal H}=-\sum_{i<j} J_{ij} \sigma_i\sigma_j - \sum_{i}h_i \sigma_i\;.
\label{eq:ham}
\end{equation}
The quenched random couplings $J_{ij}$ are independent and identically
distributed random variables taking a non zero value with a
probability decaying with the distance between spins $\sigma_i$ and
$\sigma_j$, $r_{ij}\equiv \min(|i-j|,L-|i-j|)$, as
\begin{equation}
\mathbf{P}[J_{ij}\neq 0] \propto r_{ij}^{-\rho}\quad \text{for}\;\; r_{ij}
\gg 1\;.
\label{eq:Jij}
\end{equation}
Non-zero couplings take value $\pm 1$ with equal probability.  We use
periodic boundary conditions and a $z=6$ average coordination number
\footnote{The value $z=6$ is a compromise: the computer time is
  proportional to $z$ but the critical temperature approaches zero for
  small $z$ (with drawbacks in its evaluation).}. The random field
$h_i$ is Gaussian distributed with zero average and standard deviation
$h$ \footnote{For these systems the effect of a random local field is
  very similar to a uniform one \cite{Parisi98,Katzgraber05,Young04} but has
 some advantages in performing the numerical simulations.}.  We will
denote the average over quenched disorder, both bonds and fields, by
an overline.  The universality class depends on the value of
 $\rho$.  For $\rho > 1$ it turns out to be equal to the one
of the fully connected version of the model \cite{FT_LR}, where bonds
are Gaussian distributed with zero mean and a variance depending on
the distance as ${\overline{J_{ij}^2}} \propto r_{ij}^{-\rho}$. As
$\rho$ varies, the model displays different behaviors \cite{Leuzzi08}:
for $\rho \le\rho_U\equiv 4/3$, the mean-field (MF) approximation is
exact, while for $\rho > \rho_U$, it breaks down
because of infrared divergences (IRD).  The value $\rho_{\rm U}=4/3$
corresponds to the upper critical dimension of short-range
spin-glasses in absence of an external magnetic field ($D_{\rm U}=6$).
At $\rho>\rho_{\rm L}=2$ no finite temperature transition occurs, even
for $h=0$ \cite{Campanino87}.  A relationship between $\rho$ and the
dimension $D$ of short-range models can be expressed as
$\rho=1+2/D$
which is exact at $D_{\rm U}=6$ ($\rho_{\rm U}=4/3$) and approximated
as $D<D_{\rm U}$.  Indeed, the lower critical dimension $D_{\rm L}
\simeq 2.5$ \cite{Boettcher05} corresponds to $\rho \simeq 1.8$, which
is 10\% less than $\rho_{\rm L}$.  We note that in the ferromagnetic
(ordered) Ising case on the same kind of lattices a simple theoretical
argument tells us that the value of $\rho_{\rm L}$ is $2$ for $h=0$
and $1.5$ in a field.

{\em Simulations details and data analysis --}
To study the critical behavior of the model in external field we
simulated two replicas $\sigma_i^{(1,2)}$ using the parallel tempering
(PT) algorithm \cite{PT}. Field values are $h=0,0.1,0.2,0.3$ for
$\rho=0,1.2,1.4$ and $h=0,0.1,0.15$ for $\rho=1.5$.  We used sizes up
to $L=2^{14}$ spins for $h=0$ and up to $L=2^{12}$ for $h>0$. The
number of samples is between $32000$ and $64000$ for all sizes.
Thermalization is guaranteed by the logarithmic binning (in base $2$)
of data in MC steps until at least the last two points coincide.
%%% \section{How to estimate critical temperatures}
\indent
The presence of SG long range order can be deduced from the study of
the four-point correlation function
\begin{equation}
C(x)=\sum_{i=1}^L{\overline{ \left(\langle \sigma_i\sigma_{i+x}\rangle
    - \langle\sigma_{i}\rangle\langle\sigma_{i+x} \rangle\right)^2 }}
\end{equation}
and its Fourier transform $\tC(k)$ \footnote{
In our algorithm we measure directly $\tC(k)$ and to save computing
time we express $C(x)$ as linear combination of
${\overline{\langle h_i h_j \sigma_i^{(1)}\sigma_j^{(2)}\rangle}}$,
${\overline{\langle h_i \sigma_i^{(1)}\sigma_i^{(2)}\sigma_j^{(2)}
\rangle}}$ and ${\overline{\langle\sigma_i^{(1)}\sigma_j^{(1)}
\sigma_i^{(2)} \sigma_j^{(2)}\rangle}}$.}.
Indeed, both the SG susceptibility
\begin{equation}
\chi_{\rm sg} \equiv \tC(0) 
\label{eq:TD_chi}
\end{equation}
and the so-called second-moment correlation length \cite{Caracciolo93}
\begin{equation}
\xi \equiv \frac{1}{2\sin(\pi/L)}\left[\frac{\tC(0)}{\tilde
    C(2\pi/L)}-1\right]^{\frac{1}{\rho-1}}\;
\label{eq:xiL}
\end{equation}
diverge at the critical temperature in the thermodynamic limit.
For finite (but large enough) systems, the following scaling laws hold
in the MF regime ($1<\rho\le 4/3$)
\begin{equation}
\frac{\chi_{\rm sg}}{L^{1/3}}=\tilde\chi\left(L^\frac13(T-T_c)\right),\;\;
\frac{\xi}{L^{\nu/3}}=\tilde\xi\left(L^\frac13(T-T_c)\right)
\label{eq:scalMF}
\end{equation}
with $\nu=1/(\rho-1)$, and in the IRD regime ($\rho > 4/3$)
\begin{equation}
\frac{\chi_{\rm sg}}{L^{2-\eta}}=\tilde\chi\left(L^\frac{1}{\nu}(T-T_c)\right),\;\;\;
\frac{\xi}{L}=\tilde\xi\left(L^\frac{1}{\nu}(T-T_c)\right).
\label{eq:scalIRD}
\end{equation}
with $2-\eta=\rho-1$.  Unfortunately finite size corrections to the
above scaling laws are known to be very large, especially in the
presence of an external field.  It is very important to understand
these finite size effects (FSE) and try to keep them under control.

\begin{figure}[t!]
\includegraphics[width=1.0\columnwidth]{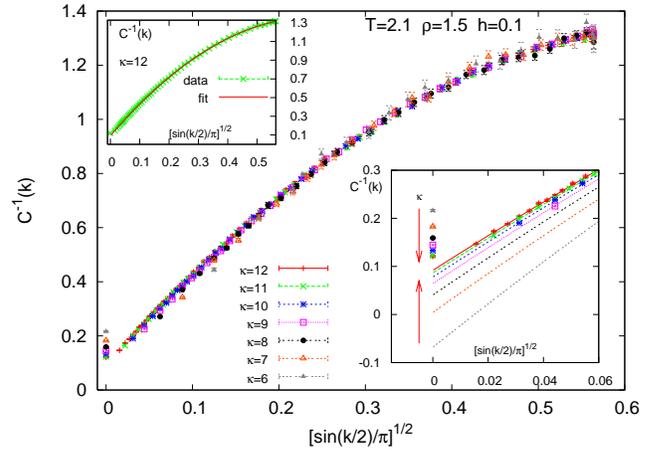}
\caption{$\tC(k)^{-1}$ vs.\ $\sqrt{\sin(k/2)/\pi}$ at $\rho=1.5$ (IRD
  regime), $h=0.1$, $T=2.1$ and $L=2^\kappa$, with
  $\kappa=6,\ldots,12$. Upper inset: quadratic fit to $L=2^{12}$ data,
  excluding $k=0$. Lower inset: comparison between $\tC(0)^{-1}$ and
  its extrapolated valued $A$.}
\label{fig:ChiInv_r15_h01_T210}
\end{figure}

In the main panel of Fig.~\ref{fig:ChiInv_r15_h01_T210} we plot
$1/\tC(k)$ versus $[\sin(k/2)/\pi]^{\rho-1}$ for an interesting case
(IRD regime with field).  The choice of the variables is dictated by
the fact that for $L\to\infty$ and $T > T_c$ the propagator on the
lattice at small wave numbers should behave like
\begin{equation}
\tC(k)^{-1} \simeq A + B [\sin(k/2)]^{\rho-1}\;,
\label{eq:propa}
\end{equation}
with $\chi_{\rm sg}=1/A$ and $\xi \propto (B/A)^{1/(\rho-1)} = (B
\chi_{\rm sg})^{1/(\rho-1)}$. In other words, $A(L=\infty,T)$ goes to
zero at $T_c$, while $B(L=\infty,T=T_c)$ stays finite.  \\ \indent We
observe in Fig.~\ref{fig:ChiInv_r15_h01_T210} that the largest FSE in
$\tC(k)$ are in $k=0$, which is the data used for estimating
$\chi_{\rm sg}$.  Moreover FSE for $k>0$ have an opposite sign with
respect to those in $k=0$ (cf. lower inset) and consequently $\xi$,
which is a function of $\tC(0)/\tC(2\pi/L)$, may be strongly affected.
The reason why FSE become smaller increasing $k$ is simple: they are
more evident in the large $x$ tail of $C(x)$ and, thus, at 
small $k$ in $\tC(k)$.  Moreover, the large $x$ part of
$C(x)$ strongly depends on $\overline{\langle q\rangle}$, which is
known to have large sample-to-sample fluctuations in a field and FSE
due to negative overlaps which should disappear in the thermodynamic
limit.
\\
\indent
With the aim of reducing FSE, we introduce a method for estimating
$T_c$ using $\tC(k)$ data with $k>0$. We fit $\tC(k)^{-1}$ by a
quadratic function $A+B y+C y^2$ with $y=[\sin(k/2)/\pi]^{\rho-1}$:
the goodness of such a fit can be appreciated in the upper inset of
Fig.~\ref{fig:ChiInv_r15_h01_T210}. As long as $T>T_c$, we expect
$\lim_{L \to \infty} A(L,T) = \chi_{\rm sg}^{-1} > 0$: the lower inset
in Fig.~\ref{fig:ChiInv_r15_h01_T210} shows size dependence of
$\tC(0)^{-1}$ and $A(L,T)$, having compatible thermodynamic limits.

\begin{figure}[t!]
\includegraphics[width=1.0\columnwidth]{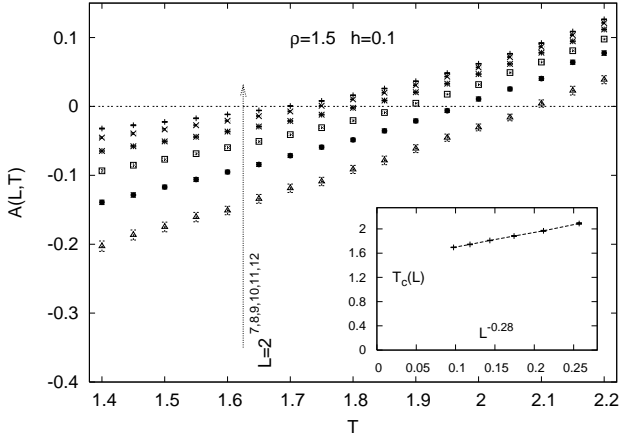}
\caption{Plot of $A(L,T)$ vs.\ $T$ at $\rho=1.5$, $h=0.1$.  Sizes are
  $L=2^\kappa$, with $\kappa=7,\ldots,12$. Inset: $T_c(L)$
  vs.\ $L^{-0.28}$.}
\label{fig:Ar15}
\end{figure}
\begin{figure}[t!]
\includegraphics[width=1.0\columnwidth]{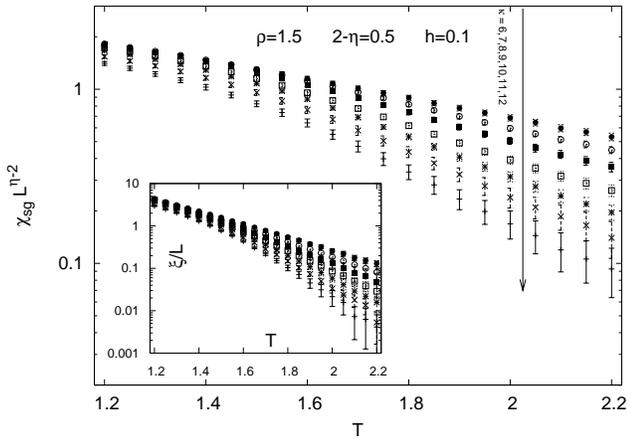}
\caption{Plot of $\chi_{\rm sg}/L^{0.5}$ vs.\ $T$ at $\rho=1.5$,
  $h=0.1$.  Sizes are $L=2^\kappa$, with $\kappa=6,\ldots,12$.  Inset:
  $\xi/L$ vs.\ $T$.}
\label{fig:cross_r15}
\end{figure}

In the main panel of Fig.~\ref{fig:Ar15} we show the best fitting
parameter $A(L,T)$ for $\rho=1.5$ and $h=0.1$.  For each size we
compute the temperature $T_c(L)$ by solving the equation
$A(L,T_c(L))=0$ (in this way only $A>0$ data are used, which are the
most reliable). Finally, we estimate $T_c= \lim_{L\to\infty} T_c(L)$
(inset of Fig.~\ref{fig:Ar15}) and obtain $T_c=1.46(3)$. The $T_c(L)$
scaling in $L^{-1/\nu}$ has an exponent $-0.28$, in good agreement
with $1/\nu=0.25(3)$ for the $h=0$ case \cite{Leuzzi08}.
On the same data ($\rho=1.5$, $h=0.1$) the analysis of the crossing
points of $\chiSG/L^{2-\eta}$ and $\xi/L$, cf. Eq. (\ref{eq:scalIRD}),
is shown in Fig.~\ref{fig:cross_r15}, yielding no evidence for a phase
transition.  The most natural explanation is the presence of
corrections to scaling laws Eq. (\ref{eq:scalIRD}).

\begin{figure}[t!]
\includegraphics[width=1.0\columnwidth]{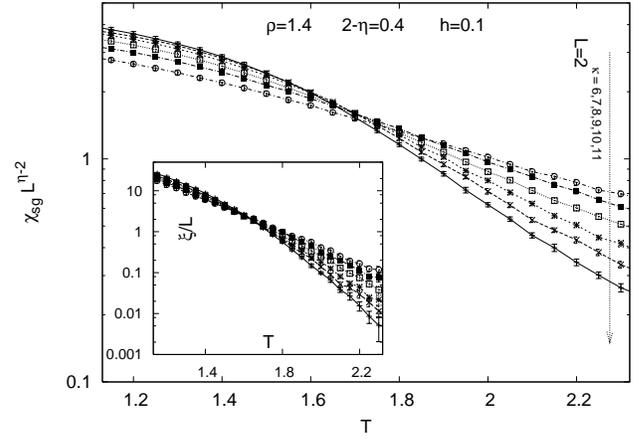}
\caption{Plot of $\chi_{\rm sg}/L^{0.4}$ vs.\ $T$ at $\rho=1.4$,
  $h=0.1$.  Sizes are $L=2^\kappa$, with $\kappa=6,\ldots,12$.  Inset:
  $\xi/L$ vs.\ $T$.}
\label{fig:cross_r14}
\end{figure}
\begin{figure}[!t]
\includegraphics[width=1.0\columnwidth]{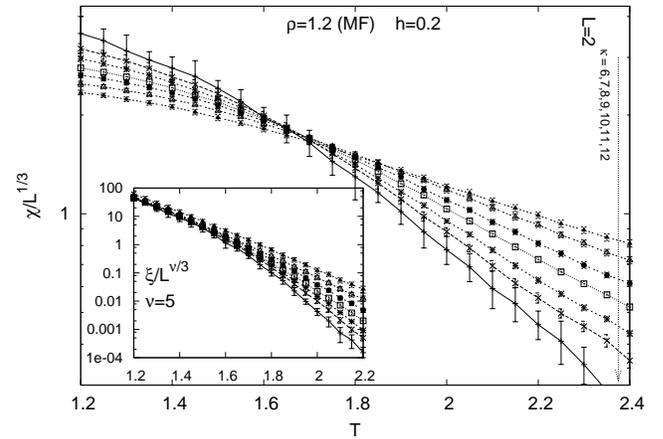}
\caption{Plot of $\chi_{\rm sg}/L^{1/3}$ vs.\ $T$ at $\rho=1.2$,
  $h=0.2$.  Sizes are $L=2^\kappa$, with $\kappa=6,\ldots,12$.  Inset:
  $\xi/L^{\nu/3}$ vs.\ $T$.}
\label{fig:FFS_Tc_rho12_chi}
\end{figure}

The case $\rho=1.4$ and $h=0.1$ provides a still more useful
comparison.  Our method of analysis returns a critical temperature
$T_c=1.67(7)$.  Fig.~\ref{fig:cross_r14} shows $\chiSG/L^{2-\eta}$ and
$\xi/L$ vs.\ $T$: crossings are present, but the curves seem to merge
for $T \lesssim 1.5$ and a precise determination of $T_c$ is
practically unfeasible.  For $\rho=1.2$, $h=.2$,
cf.~Fig.~\ref{fig:FFS_Tc_rho12_chi}, the estimate based on the scaling
of $\chi_{\rm sg}\sim L^{1/3}$ - Eq.(\ref{eq:scalMF}) - yields
$T_c=1.67(3)$, while the $\xi/L^{\nu/3}$ curves do not show any crossing
for $T>1.2$. Since the transition is known to be there in MF, this behavior of
$\xi$ is clearly caused by large FSE.

\begin{table}[ht!]
\begin{tabular}{|c||c|c|c|c|c|c}
\hline
 & $\rho$ & ``$D$'' & $h$ & $T_c$ from $\tC(0)$ & $T_c$ from $A(L,T)$\\
\hline
 & $1.2$ & 10 & $0.0$ & $2.24(1)$ & $2.34(3)$ \\
 & $1.2$ & 10 & $0.1$ & $2.02(2)$ & $1.9(2)$ \\
MF & $1.2$ & 10 & $0.2$ & $1.67(3)$ & $1.4(2)$ \\
 & $1.2$ & 10 & $0.3$ & $1.46(3)$ & $1.5(4)$ \\
 & $1.25$ & 8 & $0.0$ & $2.191(5)$ & $2.23(2)$  \\
\hline
 & $1.4$ & 5 & $0.0$ & $1.954(3)$ & $1.970(2)$\\
 & $1.4$ & 5 & $0.1$ & $\sim 1.5$ & $1.67(7)$ \\
IRD & $1.4$ & 5 & $0.2$ & $\sim 1.1$ & $1.2(2)$ \\
 & $1.5$ & 4 & $0.0$ & $1.758(4)$ & $1.770(5)$ \\
 & $1.5$ & 4 & $0.1$ & --- & $1.46(3)$ \\
 & $1.5$ & 4 & $0.15$ & --- & $1.20(7)$ \\
\hline
\end{tabular}
\protect \caption{Estimates of $T_c$: column 5 from
  Eqs.(\ref{eq:scalMF}--\ref{eq:scalIRD}) and column 6 from the
  extrapolation of $A(L,T)$ by Eq.(\ref{eq:propa}).}
\label{tab:critical}
\end{table}

Numerical values for the estimates of $T_c$ obtained with the two
methods are reported in Table~\ref{tab:critical} and look
compatible. It is clear that for large $\rho$ our method works better.
As $\rho$ is decreased, this new estimate becomes poorer, because the
scaling exponent $\rho-1$ [cf.\ Eq.(\ref{eq:propa})] is too small to
yield a robust extrapolation of $A(L,T)$.

\begin{figure}
\includegraphics[width=1.0\columnwidth]{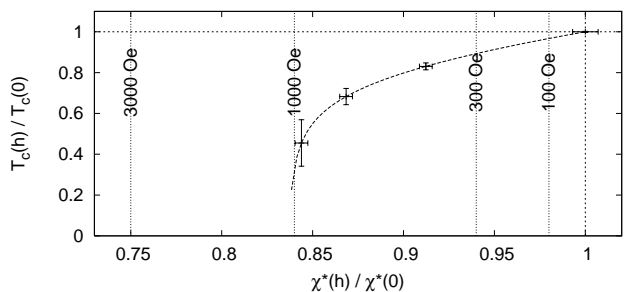}
\caption{Relative decrease of the critical temperature with increasing
  field ($\rho=1.5$, $h=0,0.1, 0.15,0.2$) versus relative decrease of
  $\chi^*$, the ZFC susceptibility at $T_c(h=0)$. The dashed curve is
  a guide for the eye.}
\label{fig:chiStar}
\end{figure}

{\em Discussion of experimental results --}
A possible objection to the presence of the SG transition (supported
by our results) is that in experiments on Ising-like SG no AT line was
detected. Here we consider, in particular, the most recent experiments
on Fe$_{0.55}$Mn$_{0.45}$TiO$_3$~\cite{Jonsson05}, where the AC
susceptibilities were very accurately measured in the presence of an
external magnetic field.  In order to relate external fields in our
model to those used in experiments we look how much the
zero-field-cooled (ZFC) susceptibility at temperature $T_c(h=0)$,
$\chi^*$, decreases as $h$ is increased.  In Fig.~\ref{fig:chiStar} we
plot $T_c(h)/T_c(0)$ versus $\chi^*(h)/\chi^*(0)$ in our model for
$\rho=1.5$.  In experiments on
Fe$_{0.55}$Mn$_{0.45}$TiO$_3$\cite{Jonsson05} with external fields of
magnitude 100 Oe, 300 Oe, 1000 Oe and 3000 Oe one has, respectively,
$\chi^*(h)/\chi^*(0)=0.98,\,0.94,\,0.84,\,0.75$. These ratios,
cf. Fig.~\ref{fig:chiStar}, suggest that a SG transition is
unlikely to be experimentally observed above $h=1000$ Oe. 

Increasing $\rho$, and/or considering $\overline{J_{ij}} \neq 0$, the
critical field decreases. The $\rho=1.5$ model considered above is
approximately equivalent to a short-range system in $D=4$. Therefore,
in order to detect, or rule out, a SG phase in $D=3$, it becomes even
more important to work at small fields.
The observation that the
fields used in experiments on Ising-like materials are maybe too large
to see a SG phase is in agreement with the results of Petit {\em et
al.}, who studied both Ising-like and Heisenberg-like spin glass
samples \cite{Petit}.

{\em Conclusions --}
In conclusion, by using a new method of data analysis, we have been
able to identify an AT transition line in the diluted power-law
decaying interaction Ising SG chain at all values of the power
analyzed, including values corresponding to short-range SG models
below the upper critical dimension.  The behaviour below the AT line
may change with the dimension. We are presently studying this
possibility.
\\
\indent
These AT lines were not found in the study of the fully connected
version performed in Ref.~\cite{Katzgraber05}, nor in
Ref.~\cite{KYultimo} where a similar diluted model was
simulated \footnote{Minor differences are present in that model: (i) a
  geometric distance in a circle, rather than a distance along the
  line with periodic boundary conditions, and (ii) Gaussian couplings
  instead of bimodal ones. These are not strongly affecting the
  critical behavior.}.  There, $T_c$ was estimated by using the
scaling properties of $\xi/L$. As we have shown, this quantity suffers
of strong FSE. We put forward an alternative method to discriminate
between a pure paramagnetic phase at all temperatures and a finite
temperature spin-glass transition.  One of the advantages of this
method is that it mainly uses data at $T>T_c$, leading to more
reliable results.
\\
\indent
For what concerns three dimensional real systems, we hint that the
magnitude of the external fields used in experiments up to now might
be too large to firmly rule out the presence of an AT transition line.
We suggest a range of fields ($h<1000$ Oe) where the transition should
take place in Fe$_{0.55}$Mn$_{0.45}$TiO$_3$ and we hope this may
stimulate further experimental investigations.
\\
\indent
This work has been partially supported by MEC, contracts
FIS2006-08533-C03 and FIS2007-60977. Part of simulations were
performed in the BIFI cluster.


\begin{thebibliography}{99}

\bibitem{Almeida78} J.R.L. de Almeida and D.J. Thouless, J. Phys. A
  {\bf 11}, 983 (1978).

\bibitem{Petit} D. Petit, L. Fruchter and I.A. Campbell,
  Phys. Rev. Lett. {\bf 83}, 5130 (1999); {\em ibid} {\bf 88}, 207206
  (2002).

\bibitem{Gabay81} M. Gabay, G. Toulouse, Phys. Rev. Lett. {\bf 47},
  201 (1981).

\bibitem{Jonsson05} P.E.~J\"onsson et al., Phys. Rev. B {\bf 71},
  180412 (2005). P.E.~J\"onsson and H. Takayama,
  J. Phys. Soc. Jpn. {\bf 74}, 1131 (2005). P.E.~J\"onsson et al.,
  J. Mag. Mag. Mat. {\bf 310}, 1494 (2007).

\bibitem{Parisi80} G. Parisi, J. Phys. A: Math. Gen. {\bf 13}, 1887
  (1980).

\bibitem{Temesvari08} T. Temesvari, arXiv:0809.1839 (2008).

\bibitem{FH88_386} D.S. Fisher, D.A. Huse, Phys. Rev. B {\bf 38}, 386
  (1988).

\bibitem{Krzakala00} F. Krzakala O.C. Martin., Phys. Rev. Lett. {\bf 85}, 3013
 (2000).

\bibitem{Marinari98} E. Marinari, C. Naitza and F. Zuliani, J. Phys A
  {\bf 31}, 6355 (1998).

\bibitem{Krzakala01} F. Krzakala et al., Phys. Rev. Lett. {\bf 87},
  197204 (2001).

\bibitem{Young04} A.P. Young and H.G. Katzgraber, Phys. Rev. Lett,
  {\bf 93}, 207203 (2004).

\bibitem{Sasaki07} M. Sasaki et al., Phys. Rev. Lett. {\bf 99}, 137202
  (2007).

\bibitem{Jorg08} T. J\"org, H.G. Katzgraber and F. Krzakala,
  Phys. Rev. Lett. {\bf 100}, 197202 (2008).

\bibitem{Leuzzi08} L. Leuzzi et al., Phys. Lett. Rev. {\bf 101},
  107203 (2008).

\bibitem{FT_LR} G. Kotliar, P.W. Anderson and D.L. Stein, Phys. Rev. B
  {\bf 27}, 602 (1983).  L. Leuzzi, J. Phys. A {\bf 32}, 1417 (1999).

\bibitem{Campanino87} M. Campanino et al., Commun. Math. Phys.  {\bf
  108}, 241 (1987).

\bibitem{Boettcher05} S. Boettcher, Phys. Rev. Lett. {\bf 95}, 197205
  (2005).

\bibitem{PT} K. Hukushima and K. Nemoto, J. Phys. Soc. Japan {\bf 65},
  1604 (1996).

\bibitem{Caracciolo93} S. Caracciolo et al., Nucl. Phys. B403, 475
  (1993)


\bibitem{Parisi98} G. Parisi, F. Ricci-Tersenghi and J. J. Ruiz-Lorenzo,
Phys. Rev. B.  {\bf 57}, 13617 (1998).

\bibitem{Katzgraber05} H.G Katzgraber and A.P. Young, Phys. Rev. B
  {\bf 72}, 184416 (2005).

\bibitem{KYultimo} H.G. Katzgraber, D. Larson and A.P. Young, Phys. Rev. Lett. {\bf 102}, 177205 (2009).
  

\end{thebibliography}
\end{document}